\documentclass[11pt]{article}

\makeatletter
\def\ps@top{\let\@mkboth\@gobbletwo
     \def\@oddhead{\rm\hfil\thepage\hfil}\def\@oddfoot{}
     \def\@evenhead{}\let\@evenfoot\@oddfoot}
\def\@bibsetup{\itemindent=-\leftmargin}
\def\@citesep{; }
\def\@cite#1#2{({#1\if@tempswa , #2\fi})}
\def\@biblabel#1{\hfill}
\def\thebibliography#1{\section*{References\markboth
 {REFERENCES}{REFERENCES}}\list
 {[\arabic{enumi}]}{\settowidth\labelwidth{[#1]}\leftmargin\labelwidth
 \advance\leftmargin\labelsep
 \usecounter{enumi}\@bibsetup}
 \def\newblock{\hskip .11em plus .33em minus -.07em}
 \sloppy
 \sfcode`\.=1000\relax}
\renewcommand{\section}{\@startsection {section}{1}{\z@}{-3.5ex plus -1ex minus 
    -.2ex}{2.3ex plus .2ex}{\centering\large\bf}}
\renewcommand{\subsection}{\@startsection{subsection}{2}{\z@}{-3.25ex plus
    -1ex minus -.2ex}{1.5ex plus .2ex}{\centering\bf}}
\makeatother
\begin{document}
\mbox{}
\vspace{1in}
\begin{center}
{\Large\bf A High Spectral Resolution Atlas of Comet 122P/de Vico} \\ [25pt]
Anita L. Cochran and William D. Cochran \\
Department of Astronomy and McDonald Observatory \\
University of Texas at Austin, Austin, TX \\
Accepted for publication in {\it Icarus} \\
\end{center}
\vspace{1in}

\begin{center}{Abstract}\end{center}
On 3 and 4 October, 1995, we obtained high spectral resolving power
(R$=\lambda/\Delta\lambda = 60,000$) observations of comet
122P/de~Vico using the 2DCoude cross-dispersed echelle
spectrograph on the 2.7-m telescope of McDonald Observatory. 
The spectra cover the wavelength range from 3830\AA--10192\AA.
The spectra from 3830--5776\AA\ are continuous;
from 5777--10192\AA\ there are increasing interorder gaps.
The comet was at a heliocentric distance
of 0.66\,{\sc au} and a geocentric distance of 1.0\,{\sc au}.
Comet de~Vico has a very high gas-to-dust ratio and the spectra
have excellent signal/noise.  These two factors combined to yield
spectra with a large number of emission lines.
We have collected laboratory molecular line lists
and have used these line lists in order to identify as many of the
detected lines as possible.  We have identified 12,219 emission lines
and have located another 4,055 lines which we cannot identify.
We present representative spectra and identifications along
with a description of our plans to make these spectra and identifications
available to the community.  This atlas should prove a valuable
tool for future studies of comets.


\newpage
\section{Introduction}

Understanding the physics and chemistry of comets can lead to significant
improvements in our knowledge of
the formation of the Solar System.  Such studies allow constraints on
the physical conditions in the solar nebula, on the dynamical evolution
of the disk and of planetesimals, and on the chemical evolution of these
bodies.  
High resolving power spectroscopic observations offer unique
opportunities for studies
of the origin and subsequent chemistry of comets.  
For example, measurement of atomic isotope
ratios in a sample of ``Kuiper belt'' and ``Oort cloud'' comets
can place valuable new constraints on physical processes in the
solar nebula and proto-planetary disk during the comet formation
epoch.

The complex molecular spectrum of the coma of a comet results
in thousands of lines being visible in the spectral region
from 3800--10200\AA.  In order to undertake studies of a particular
feature or molecule, one must be able to identify most of the
lines in the region of interest.  Yet many spectral regions are
complicated by the superposition of lines from more than one
molecule.  In addition, laboratory studies of diatomic and
polyatomic molecules are often difficult to obtain and incomplete.
Valk {\it et al.} (1992) published a catalogue of line identifications
for the near-UV and blue region of the spectrum of Comet C/1989\,X1 (Austin)
from spectra at moderate spectral resolving power (R=2,500).
Brown {\it et al.} (1996) published a catalog
of cometary emission lines from comets 109P/Swift-Tuttle and 
C/1989\,X1 (Austin).
They found 2997 lines and were able to identify 2438 of them.
Zhang {\it et al.} (2001) utilized high resolution spectra of 
C/1995\,O1 (Hale-Bopp)
and the catalogue of Brown {\it et al.} in order to identify
532 emission lines (73 unidentified).
These catalogues are a critical first step for cometary science.

This paper represents an update to the catalogues of Brown {\it et al.}
and Zhang {\it et al.}	
We present observations of comet 122P/de~Vico obtained 
at McDonald Observatory in October 1995. 
Comet de~Vico is the perfect target for such
a study since it has a very high gas-to-dust ratio and was reasonably
bright, allowing for spectra of exceptionally high signal/noise with
very little continuum (signal/noise at peak was $>$300).
We identified 12,219 lines in the spectrum of de Vico.  In addition,
there are 4,055 unidentified lines in our spectra.  

\section{Observations and Reductions}
We observed comet 122P/de Vico with the
2.7-m Harlan Smith telescope of McDonald Observatory and the
2DCoude spectrograph (Tull {\it et al.} 1995) during October 1995 (see
table~\ref{log} for information on the observations).  These
observations were obtained at a spectral resolving power of 60,000
and covered the wavelength range from 3800--5776\AA\ continuously, with
spectra with increasing interorder gaps from
5776--10192\AA.
Table~\ref{wave} indicates the spectral coverage of each of
the echelle orders.  In all cases, the slit was 1.2 arcsec wide
by 8.2 arcsec long.
The spectra were extracted using variance weighting along the
spatial dimension, with no attempt made to preserve
spatial information.  The signal/noise
was extremely high throughout.  De Vico represents a perfect
comet for such identification work since the gas/dust ratio is
one of the highest known for a comet.  This results in spectra
with almost no continuum so that even weak features are unambiguously
detected.  In addition to spectra on the optocenter, we obtained
one spectrum with the telescope offset 100\,arcsec into the tail.
These tail spectra are useful for distinguishing between neutral and ionized
species.

Incandescent lamp spectra were obtained in order to flat field
the data.
The wavelength calibration was made by fitting a two-dimensional
function to observations of a ThAr hollow-cathode lamp.
Over 3100 lines were fit, with a fourth order fit along an order
and fifth order fit across orders, resulting in an rms error 
of the wavelength of $\sim$2.5m\AA.
This is comparable to, or better than, typical laboratory spectra described
in this paper.
The cometary spectra were Doppler-shifted by a velocity equal to the
geocentric radial velocity of the comet (i.e by $-\dot\Delta$ from table~\ref{log}
to bring them onto the laboratory rest frame).

A cometary spectrum consists of molecular and atomic emission lines,
superimposed on a continuum which results from sunlight reflecting
off the dust. 
It is necessary to remove the underlying continuum from the emission
spectrum if the goal of
a study of the spectra is to determine the amount of a particular gas.
However, removal of the continuum is fraught with uncertainties in
the color of the dust and in photon statistics.  Thus, removal of the continuum
can induce noise into spectra.
Fortunately, comet de~Vico has a high gas/dust and
the continuum is essentially negligible at all wavelengths. 
Given this and the problems associated with continuum removal,
we chose \underline{not} to 
remove the continuum for our purposes in this paper of line identification.
This results in an offset of the spectrum from zero (typically the
continuum for de~Vico was about 350 counts and the emission lines
are many thousands of counts).  The continuum can also result in the apparent diminution
of the strengths of emission lines when they coincide with a solar absorption
feature.  This might result in our not detecting a
very weak feature.

In addition to the underlying continuum spectrum, there is a superposition
of a telluric spectrum onto the cometary spectrum.  This spectrum
consists of emissions due to metastable oxygen, sodium, OH and O$_2$
and absorption due to O$_2$ and H$_{2}$O.  The telluric spectrum
is at the rest frame of the Earth's atmosphere and is, therefore,
shifted by $-\dot\Delta$ with respect to the cometary rest frame.
Thus, we Doppler shifted by this same velocity the telluric emission line lists discussed
in the next section to identify the telluric features present in the Doppler-shifted 
cometary spectra.

The telluric emission lines are generally narrow and only affect the
wavelength at which they occur.  The telluric absorption is a different
problem entirely.  In the regions of the strong
absorptions, such as the O$_2$ ``A" band or the 9300\AA\ H$_{2}$O band,
almost all photons entering the atmosphere are absorbed
before they reach the telescope.  Thus, even if a line at these
wavelengths is excited in the cometary spectrum, we will not
record the feature in our spectrograph.  There is no way to remove
the effect of these saturated telluric absorptions and the information
from the comet
is forever lost (unless a spectrum taken at a different time has
a Doppler shift that moves the cometary line and the telluric absorption
apart).  The orders most affected by the telluric
absorption are orders 50 (O$_2$ ``B" band), 48 and 47 (H$_{2}$O ``a"
band), 45 (O$_2$ ``A" band), 42 and 41 (H$_{2}$O ``z" band), and
38--35 (H$_{2}$O ``$\gamma$" and ``$\rho$" bands).

\section{Linelists}
One of the most difficult aspects of compiling a cometary spectral atlas
is finding good laboratory line lists of all of the relevant molecules. 
As a first step in building our spectral atlas, we
searched out many molecular line lists. 
Many of
the molecules of importance were studied years ago and the
line lists do not exist in digital form; we have spent many hours
scanning line lists and typing in and checking large quantities of numbers
by hand 
(possibly with undetected errors introduced).  Some of the relevant
molecules are not well understood and there are many unassigned
lines. 
For some molecules, we
have used work by several different authors to be as complete
as possible.  In the case of some molecules, such as CN, we had
access to line lists which were much more inclusive than necessary.
The tables of the identified molecular lines which we include as
part of this paper list only those lines which we believe we detected.
However, such over-complete line lists may prove useful for future
cometary spectra.  We note, also, that in the red and near-IR, there
are interorder gaps in our data.  In these regions, we leave gaps
in the lists of identified lines.  Refer to table~\ref{wave}
for a complete list of the wavelengths for which we have
observations.

\subsection{C$_{2}$}

There are two principal band systems of C$_{2}$ which are observed in
the optical in the spectra of comets.
These are the Swan, or d\,$^3\Pi_g$~--~a\,$^3\Pi_u$, system, and
the Phillips, or A\,$^1\Pi_u$~--~X\,$^1\Sigma^+_g$, system.
The Swan system is dominant in the green, orange and red region of the 
spectrum, while the Phillips is important in the near-IR and IR.

For the Swan system, we used the line list of Phillips and Davis (1968).
We extended this line list for high J-values of the 0--0, 0--1 and
2--1 bands using the wavelength spacing of the highest previously defined
lines to identify which emission lines were the continuation of these bands.
We derived slightly different wavelengths for four of the highest J-value lines of
the R-branches of the 1--0
band than in the original reference ($\Delta\lambda = $0.042-0.076\AA).  
Most of our extensions are
in agreement with the additional wavelengths also derived by 
Brown {\it et al.} (1996), though we observed to higher J-levels
for the 0--0 band. 
 
For the Phillips system, we used the works of
Ballik and Ramsay (1967) and Chauville {\it et al.} (1977).

\subsection{C$_{3}$}

The C$_{3}$ molecule was first observed in cometary spectra in 1882,
although the species causing the emission feature in comets was
unknown.
It is difficult to study in the laboratory because
it is relatively unstable.  Thus, the molecule causing the
emission lines was not identified until 1951.
In cometary spectra, the important bands are known as the
``4050-\AA\ Group", indicative of the peak of the spectrum.
However, careful inspection of cometary spectra show that lines
attributable to this band can be seen from approximately
3350\AA\ to 4700\AA, although the main part of the
band is between 3900 and 4140\AA\ with a maximum at 4050\AA\ and
an additional (but not well identified) peak at 4300\AA.

The structure of this molecular band is quite complex since 
the bending frequency of the lower state is small and there is
a large Renner-Teller splitting.  The band has been identified
as a A\,$^1\Pi_u$ -- X\,$^1\Sigma_g^+$ electronic transition
(Gausset {\it et al.}, 1965). R-branch bandheads at 4072\AA\ and
4260\AA\ can be assigned to the (1,0,0)--(1,0,0) and (0,0,0)--(1,0,0)
bands respectively
(Merer, 1967).   The density of the lines results
in a pseudocontinuum from C$_{3}$.

C$_{3}$ is a linear, symmetric molecule containing nuclei which are
identical in mass.  For all $\Sigma$ states of the molecule, every
second line is missing.  In other states, one member of the $e/f$-parity
doublet of each rotational level is missing, although all rotational
levels are represented (Tokaryk and Chomiak, 1997).

Most of our line identifications come from Gausset {\it et al.} (1965),
with additional identifications from Merer (1967) and 
Tokaryk and Chomiak (1997).  Tokaryk and Chomiak argue convincingly that the
(0,2,0)--(0,0,0) and (0,2,0)--(0,2,0) rotational assignments of
Gausset {\it et al.} 
are in error.  Thus, in the wavelength region from 3990--4019\AA\ we
have preferentially used the Tokaryk and Chomiak assignments.
However, Tokaryk and Chomiak do not report rotational assignments
for these bands outside this wavelength region, so we have continued
to use the Gausset {\it et al.} assignments for these two bands
at other wavelengths. 

The combination of the complexity of the spectrum and the difficulty
of studying this molecule in the laboratory has meant that 
the line identifications are far from complete and may
not always be accurate.  This is readily
apparent at 3949\AA\ where we observe an obvious C$_{3}$
bandhead which is not in the line lists cited here.
The lines which \underline{are} identified in this region arise from
the (0,4,0) -- (0,2,0)
band Q-branch, but the line identifications of Gausset {\it et al.} do
not include any assignments for the R- or P-branches.
Balfour {\it et al.} (1994) have studied C$_{3}$ at these wavelengths
and do indeed identify this bandhead as the R-branch of the
(0,4,0) -- (0,2,0) band and identify other lines in the region as from the P-branch.
However, the Q-branch assignments of Balfour {\it et al.} disagree
with those of Gausset {\it et al.} Tokaryk and Chomiak have pointed
out that the Balfour {\it et al.} spectra are at much lower resolution
than the other spectra and that Balfour {\it et al.} may not have been able to
disentangle the  superimposed bands.  We have no basis for choosing
between the various assignments.  We have chosen to neglect the Balfour
{\it et al.} assignments entirely based on the arguments of Tokaryk and
Chomiak.  However, it is reasonably certain that the undentified lines near
3949\AA\ are from the R-band of the (0,4,0) -- (0,2,0) band.

We do not resolve cleanly the C$_{3}$ lines in our spectra. 
The outflow of the gas in the coma broadens the line sufficiently
that even at $3\times$ the resolution with R=180,000,
a resolving power we used with
comets Hyakutake and Hale-Bopp, the C$_{3}$ is not resolved in the
densest regions.
Thus, we have not always been able to definitively match
a line list with our spectra.  In the densest regions, we have
just assumed that all known lines were observed.  There are
some known lines (e.g. in the (0,0,0)--(0,4,0) or (0,0,0)--0,6,0) bands) which
we definitely {\it do not} observe in our spectra.


\subsection{CH}

A wonderful new, inclusive line list has recently become available for
studies of CH.  This is the SCAN-CH list of J{\o}rgensen {\it et al.} (1996).
This list includes transitions of $^{12}$C$^1$H and $^{13}$C$^1$H
for the infrared transition X\,$^2\Pi$ -- X\,$^2\Pi$ and
three electronic transitions A\,$^2\Delta$ -- X\,$^2\Pi$,
B\,$^2\Sigma^-$ -- X\,$^2\Pi$, and C\,$^2\Sigma^+$ -- X\,$^2\Pi$.
A total of 112,821 lines are given for each isotope!
We identified low-J lines of the 0--0 band of the B--X transition
around 3900\AA\ (some of which can be confused with isotopic lines
of CN) and low-J lines of the 0--0 and 1--1 bands of the
A--X transition around 4300\AA.  This new SCAN-CH line list includes
the satellite transitions of the CH bands, which are observed
in our spectra.

\subsection{CN}

Great effort has been expended to understand the CN molecule
because it is a significant source of opacity in some stars.
Thus, there are CN line lists which are much more inclusive than
what is needed for cometary studies, where depopulation of the
upper J-levels is relatively rapid.

Two different electronic band systems are seen in cometary spectra,
the so-called violet system, or the B\,$^2\Sigma^+$ -- X\,$^2\Sigma^+$
band, and the so-called red system, or the A\,$^2\Pi$ -- X\,$^2\Sigma^+$
band. 

For the violet system, we used the line list of
Kurucz (1995, CDRom 18) which contains about 350,000 lines.
Lines of both $^{12}$C$^{14}$N and $^{13}$C$^{14}$N are included 
with J up to 150.5.
For our purposes, we have only used the $^{12}$C$^{14}$N wavelengths.
Indeed, some of our ``unidentified" lines may be lines
of $^{13}$C$^{14}$N.
Since this band is a $\Sigma$--$\Sigma$ transition, there are P- and R-branches
but no Q-branch.

For the red system, we used the line list of Davis and Phillips (1963).
J{\o}rgensen and Larsson (1990) recently calculated  CN red line lists
with up to 1 million lines for $^{12}$C$^{14}$N.  However, the
J{\o}rgensen amd Larsson work is most concerned with completeness for
line blanketing and is not derived from measured line positions.  Thus,
some of the line center uncertainties in this list are higher than we preferred,
which is why we used the older reference.

\subsection{NH$_{2}$}

Throughout much of the red region of a cometary spectrum, there
are scattered emission lines which are attributable to NH$_{2}$.
The transitions responsible for these lines are the
\~A$^2$\,A$_1$ -- \~X$^2$\,B$_1$ electronic transition along
with transitions between the high vibronic levels and the ground
state of the \~X$^2$\,B$_1$ electronic band (e.g.
(0,13,0)\~X$^2$\,B$_1$ -- (0,0,0)\~X$^2$\,B$_1$).
The spectrum of the NH$_{2}$ band is quite complex and irregular
because the upper level is linear and the lower level is bent at an angle
of 103$^\circ$.
Also, perturbations are important, as is the Renner-Teller effect.

The traditional reference for cometary NH$_{2}$ line lists
is Dressler and Ramsay (1959).  While this reference is
moderately complete, it has been superseded in subsequent
years by several other laboratory analyses.
Most important in newer studies is the inclusion of more complete
perturbation models and higher quality laboratory spectra.
For our line lists, we have used the data of 
Ross {\it et al.} (1988) for the wavelength range 5400 -- 6800\AA.
Redward of 7000\AA\ we used the data of Johns {\it et al.} (1976)
and of Huet {\it et al.} (1996), with updates by Hadj Bachir {\it et al.} (1999).
From 6800--7000\AA\ (the (0,2,0)--(0,0,0) and (0,1,0)--(0,0,0) bands),
the data of
Dressler and Ramsay and Johns {\it et al.} (1976) are
most complete,
so we use these two lists preferentially in that region.
For regions blueward of 5400\AA, the Dressler and Ramsay 
wavelengths are really the only data in existence; we used
their wavelengths in this blue region.

Since NH$_{2}$ has the perturbations and the bent state, modern
works prefer to use the bent notation for the vibronic structure.  This
is in contrast with the work of Dressler and Ramsay and Johns {\it et al.}, in
which
the vibronic states were denoted in the linear notation.
In our line lists, we have converted all of the 
vibronic notation from the linear to the bent notation.
In the linear notation, sub-bands were denoted by $\Sigma$, $\Pi$,
$\Delta$, $\Phi$, and $\Gamma$ (for K$=0,\pm1,\pm2,\pm3,\pm4$ respectively).
To convert from a linear to a bent notation, one uses the formula
\[ v^{lin} = 2 v^{bent} + |l| \]
where $l = \rm{K} \pm \Lambda$.  Thus, the old notation for
the band around 6500\AA\ would have been (0,8,0) $\Pi$ or
(0,8,0) $\Phi$, while now those bands are denoted (0,3,0) and
(0,2,0) respectively [in these examples the lower level is
assumed to be (0,0,0)].  
Despite the fact that the linear notation has been exclusively in use
by the comet community, we feel it is more appropriate to adopt the
same notation as the physicists who measure these molecules.

\subsection{CH$^+$}

The CH$^+$ lines which are seen in cometary spectra are from
a $A\,^1\Pi - X\,^1\Sigma^+$ transition which contains P, Q and R branches.
The line list used for these bands was from Douglas and Herzberg (1942).
The bands degrade towards the red.
We only see lines of the 0--0 branch with relatively low J-values.
While we see very few CH$^+$ lines in our spectra, they are visible
in both the optocenter and the tail spectra.

\subsection{CO$^+$}

There are three band systems which can be attributed to the CO$^+$ molecule.
These are the first negative carbon bands (B\,$\Sigma$ -- X\,$^2\Sigma^+$),
which occur in the UV, the comet-tail bands (A\,$^2\Pi$ -- X\,$^2\Sigma^+$),
occurring in the violet through red, and the Baldet-Johnson bands
(B\,$^2\Sigma^+$ -- A\,$^2\Pi$) also in the violet.  Observations
in the optical bandpass do not include the first negative carbon bands
(however, UV observers can find wavelengths for these lines in
Rao (1950a).  Rao (1950b) showed that previous optical band assignments 
were in error and that the Baldet-Johnson system was much weaker
than the comet-tail bands.  Thus, we adopted the band assignments
of Rao (1950b) for the comet-tail bands and neglected the
Baldet-Johnson system entirely.

For our comet-tail line list, we used four references:
Coster {\it et al.} (1932), Schmid and Ger{\"o} (1933), 
Rao (1950b) and Haridass {\it et al.} (2000). 
We updated the band designations of the first
two references with the designations in Table 2 of Rao and also
corrected the values for N in these papers as outlined by Rao.
Unlike for CH$^+$ and H$_{2}$O$^+$, the CO$^+$ lines could not be
detected in the optocenter spectra.  These were entirely found in
the tail spectrum.  Indeed, we detected only the 2--0 band of CO$^+$
in the de~Vico tail spectrum.

It is unusual to detect the 2--0 band without the 3--0 band since the 3--0 band
is normally the strongest CO$^+$ band in cometary spectra.  The ion tail of de~Vico
was generally quite narrow so it is possible that our ``tail" spectrum missed the
strongest emissions, accounting for the weakness of the CO$^+$ emissions.

\subsection{H$_{2}$O$^+$}

With H$_{2}$O being the dominant ice in the cometary nucleus,
it is not surprising to see H$_{2}$O$^+$ emission in the ion
tail spectrum of a comet.  
The electronic transition is \~A$^2$\,A$_1$ - \~X$^2$\,B$_1$,
which is observed from
4000--7500\AA.  Much of the structure
of the H$_{2}$O$^+$ molecule is similar to that of the NH$_{2}$
molecule, including the strong Renner-Teller effect.

The most complete reference for the wavelengths of the H$_{2}$O$^+$
lines is Lew (1976).  In this paper, the bands are denoted
in the linear notation (see the section on NH$_{2}$) whereas
the bent notation is considered correct by current standards.
Therefore, we have converted the H$_{2}$O$^+$ bands to the
bent notation in a manner which is identical to the NH$_{2}$.

As with the CH$^+$, the H$_{2}$O$^+$ lines are apparent in both
the optocenter and the tail spectra.  We required that a line be
visible in the tail spectrum before we attributed it to H$_{2}$O$^+$.

\subsection{Night Sky Emission Lines}
Finally, even at the high spectral resolving power of our spectra,
coupled with the small entrance slit, one does end up observing
emission lines produced in the Earth's atmosphere.
These lines must be accounted for when identifying lines.
Fortunately, a high-resolution atlas of night sky
emission lines has recently been compiled by Osterbrock {\it et al.} (1996)
for just this purpose.  The dominant molecular emission comes
from OH, with contributions of O$_2$ and some atomic lines.
We have applied a Doppler shift equal to $-\dot\Delta$ to
the Osterbrock {\it et al.} wavelengths
to compensate for correcting the cometary spectra to the laboratory
rest frame.  We confirmed that the correct Doppler shift was applied
by checking the strong night sky atomic lines (Na D, O ($^1$D) and
H $\alpha$).

In the lists of identified lines which follow, we do not include the
night sky lines, though
they are represented on our plots.  We have drawn no conclusions about
their existence in our spectra, assuming they are all present.
Instead, an otherwise unidentified line which is coincident with
a night sky line is considered to be that night sky line without
taking into account the line's strength.  Some unidentified cometary and night
sky lines could, however, be coincident.

\subsection{Additional Molecules}

The molecules which have been identified above are the same ones which
Brown {\it et al.} (1996) found in their spectra and which previous
authors have identified.  Brown {\it et al.} also investigated a
range of other molecules which did not show up in their
spectra and we have assumed they are not in our spectra either.

There are some potential molecules which we checked but our investigations
proved inconclusive.  One is H$_{2}$O, which is the dominant
molecule in cometary nuclei and which has many transitions in
our spectral region.  The wavelengths of these lines are well known.
However, though we started out examining our spectra for H$_{2}$O lines,
we quickly realized that it would be implausible that we could
detect these lines in our spectra when the Earth's
atmosphere is so full of H$_{2}$O.  

Another potential molecule in cometary spectra is N$_2$.  Again, the
atmosphere of the Earth makes this molecule impossible to study,
so we did not search for it.

One molecule for which we searched was N$_2^+$. 
Cochran {\it et al.} (2000) showed
that this molecule was not present in spectra of de~Vico or
Hale-Bopp and placed stringent upper limits on its presence.

\section{Line Identifications}

Once the molecular line lists were compiled, it was necessary to
examine all of the spectral orders and determine which lines we
detected.  First, we scrutinized  our spectra
for wavelength coincidences of the emission features and 
lines in the line lists.
The strongest lines are quite easy to see in the spectra, but the weakest
lines needed verification.  We utilized spectra from both nights
to test the veracity of a detection of a weak line.  We required
that the line be in both night's spectra before we would categorize
the feature as a detection.  Our very high signal/noise, coupled
with having multiple spectra, allowed for the detection of some
very weak features.

With the density of the lines in some of these bands and in
our spectra, it is expected that there will be some coincidences
which are accidental.  
When a line from the line list was coincident with an emission line,
we checked to determine whether detection of the line made sense.
This criterion took the form of requiring that the lines originating
from lower energy levels
be observed before believing that lines from higher energy transitions were
detected.
This is equivalent to assuming that the upper level populations can be
described by some sort of effective Boltzmann temperature, which is not
necessarily a physical temperature. 
Indeed, because C$_{2}$ is a homonuclear molecule, with no allowed dipole
transition, high-J levels can be quite strong because these
levels get ``pumped up". 
This criterion was applied on a band-by-band basis.
Thus, there were times when an ``obvious" detection was deemed
accidental because lines with lower excitation energies for the
same band were not seen.  We would then eliminate these accidental
detections from our list.

Conversely, there are some times when we included a line, though it
is not very obvious, because it is more favorable energetically than
lines which were detected.  A possible cause for the absence of an
expected line is coincidence with a solar absorption feature.

For the ions, we checked the tail spectrum as well as the optocenter
spectra in order to determine if a line was ionic in nature.  
A simple coincidence of the ionic species wavelength to a feature in
the optocenter
spectra was not considered sufficient evidence that a line was an ion.
The line also had to be present in the tail spectrum.  As noted
above, the CH$^+$ and H$_{2}$O$^+$ lines were observed in both
the optocenter and tail spectra, but the CO$^+$ was only observed in
the tail spectrum.  It was considered sufficient to identify an ionic
line if it occurred solely in the tail spectrum.

We compared our tail spectrum of de~Vico with the list of identified
and unidentified
species in the spectrum of the tail of comet C/1996\,B2 (Hyakutake)
(Wyckoff {\it et al.} 1999).
Some of Wyckoff {\it et al.'s} H$_{2}$O$^+$ lines are seen in our
spectra, while others
are not, and we detect many H$_{2}$O$^+$ lines that they did not see
(however, our data are
at much higher spectral resolution and so exact matches in wavelength
with their identifications are difficult).
We cannot confirm most of their lines; though a few
of their lines other than H$_{2}$O$^+$ might be coincident
with features in the de~Vico tail spectrum, these features are generally also
coincident with neutral lines identified in the optocenter spectra.
The Hyakutake tail spectrum of Wyckoff {\it et al.} was obtained at
larger distances from the
nucleus than was our de~Vico tail spectrum, which may explain why
we do not see the same lines.  
In addition, with the high airmass of observations and a relatively small
slit on our instrument, we could have missed the densest part of the ion
tail (we do see some ions though).
However, the explanation could as
easily be that de~Vico does not have a very strong tail spectrum.

In the blue part of the spectrum, there was an increasing wavelength
overlap between the echelle orders as the wavelength decreased
(higher order number). 
This allowed for the comparison
of many lines in two different echelle orders.  In general, we found the
agreement excellent, with wavelength matches of $\sim0.01$\,\AA\ or
better.  However, for the bluest orders, the spectral overlap
regions were falling well
off the blaze of the grating and, consequently, became noisy in the bluest
quarter of these orders.  Therefore,
we gave priority for identification of lines in the overlap regions to
the red end of one order over the blue end of the next lowest (and therefore
redder) numbered order.

One interesting result of our identifications comes from the C$_{2}$ Swan
bands, particularly the $\Delta v = -1,\,0,\,{\rm and}\,1$ bands.
We observe lines with very high J-values, up to J=109!
However, the lowest few J-value lines are missing or very weak.
This is probably indicative of two different rotational temperatures
for C$_{2}$, though we have not modeled these lines yet to demonstrate
this.  Such an effect was also seen by Lambert {\it et al.} (1990)
for comet Halley.

For the CN red system, we found that for a particular J level, the
P$_{12}$ and R$_{12}$ lines were generally quite weak when compared with
the P, R or Q lines of the same J level.  Inspection of the
line strengths in the laboratory work of Davis and Phillips (1963) shows that
this is generally seen for this molecule and is normal.

In addition to the molecular emission lines, we found three atomic
emission lines in the spectrum.  These are the red metastable oxygen
doublet, O ($^1$D) 6300\AA\ and 6364\AA, and H $\alpha$.
The green oxygen line at 5577\AA\ is cloaked by the 1--2 C$_{2}$ emission
band; the Na ``D" doublet is in the wavelength region of the C$_{2}$ $\Delta v=-2$
bands.  However, at most, there is just a weak hint of the cometary Na features at 
appropriate wavelengths.  The weakness of any Na emission in de~Vico is quite 
unusual in a comet at de~Vico's heliocentric distance.

Finally, when we had exhausted our line lists, we still had a great many
emission lines in our spectra with no identifications.  We have flagged
these in our line lists with the tag ``Unid" next to the measured
wavelengths and computed vacuum frequencies. 
A few of these ``unidentified" lines are probably isotopic lines,
e.g. $^{13}$CN vs. $^{12}$CN.  Isotopic lines are quite weak but
we do know of several in our spectra.  However, we have chosen
to ignore them for now, because proper identification takes modeling
the band in question, which is beyond the scope of this atlas.

Figure~\ref{order49} shows a plot of the blue half of order 49,
covering the wavelength range from 6957--7014\AA.
In this figure, we show spectra from both nights, offset from one
another for clarity.  The spectra are plotted with an expanded y axis
in order to illustrate the weaker features.  The spectra are
plotted in three panels to allow for inspection of the features.
Plotted below the spectra are tick marks for all of the identified
lines, in this case lines of CN, C$_{2}$, NH$_{2}$ and H$_{2}$O$^+$.
Plotted above the spectra are tick marks for the night sky OH (Doppler
shifted to the appropriate rest frame for the comet) and the position
of lines we could not identify.
Dashed vertical lines mark the position of the features in the spectra.
Examination of Fig.~\ref{order49} shows that there are a substantial number
of quite strong lines which are unidentified, including lines with more
than 20,000 counts!

In contrast, Figure~\ref{order77} shows order 77 from 4 October.
There are obvious emission features (and solar absorptions) in this
spectrum.  Indeed, we have located 64 lines in this spectral region
which are visible in spectra from both nights.  However, we can
not identify the molecule for \underline{any} of these 64 features.
This order is an excellent example of the desperate need for more
laboratory studies.

Figure~\ref{wavel} shows a plot of all of our detected lines, by molecule,
as a function of wavelength. Unidentified features are also marked.
The gaps which can be seen in the red
for molecules such as C$_{2}$ and CN are the result of the interorder
gaps in our spectra.  Molecules such as C$_{2}$ and NH$_{2}$ can be
found at almost any optical wavelength, while C$_{3}$ and CH are
confined to the blue.  

Table~\ref{lines} gives a representative listing of the identifications for
the wavelength region from 6957--6976\AA, the spectral region
shown in the top panel of Fig.~\ref{order49}.  
A table similar to Table~\ref{lines} for each of
the molecules
shown in Figure~\ref{wavel} is available in digital (ASCII text) 
form\footnote{These tables and the plots mentioned below may be obtained
from the web site (currently \mbox{pdssbn.astro.umd.edu}) of
the Small Bodies Node of the NASA Planetary Data System}.
These tables contain the wavelength and vacuum frequency of the lines,
along with the band designations and transition.
A table of the unidentified lines is also available in digital form. 
We also will make available a table of all of the identified and
unidentified features, sorted by wavelength.  

As with Table~\ref{lines}, the digital tables will be entirely in
ASCII characters and will not use any sub-
or super-scripts; they can be utilized on any computer without having to deal
with embedded special characters.  The tables are in fixed format.
For most molecules, the electronic transitions are listed by
their letter designations (A, B, X) instead of their term designations.
For C$_{2}$, we use the more common names of Swan and Phillips,
where the Swan band is the d\,$^3\Pi_g$~--~a\,$^3\Pi_u$ band
and the Phillips is the  A\,$^1\Pi_u$~--~X\,$^1\Sigma^+_g$ band.
As vibrational transitions carry 
designations with different forms for diatomic and polyatomic
molecules, those transitions are listed in the manner appropriate
for each molecule.    In all cases, the upper state is listed first, then
the lower state.  For the rotational transitions, the designations
are very dependent on the individual molecule, since some just
have P, Q and R branches, some, such as CN A-X, have branches such as
P$_{12}$, and some, such as CH, have branches such as $^R$Q$_{21}fe$
(which is listed as ``$^\wedge$RQ21fe" in plain ASCII).
Molecules such as NH$_{2}$ and H$_{2}$O$^+$ have rotational
states which are designated with three quantum numbers of the form
$N^{\prime}_{K^{\prime}_{a}K^{\prime}_{c}} - N^{\prime\prime}_{K^{\prime\prime}_{a}K^{\prime\prime}_{c}}$.
The NH$_{2}$ and H$_{2}$O$^+$ rotational states are separated into F$_1$
and F$_2$ components.  We have not included those designations
in our tables.

Sometimes more than one transition occurs at the same frequency.
We list all rotational transitions from the same vibrational and electronic
band on the same line in the tables.  When there are concurrent transitions
from different vibrational bands, we list the two different band
transitions on separate lines of the table.

We chose to list only the wavelength and frequency
of the unidentified lines and not their strengths. 
There were several reasons behind
this decision.  First, we have no absolute flux calibration of
these coud\'{e} spectra so there is no
way to calibrate the blaze function.  Additionally, we have not
attempted to remove the atmospheric extinction and the comet
was observed at an airmass of 2--3 airmasses.  This means that
relative strengths within an order are probably accurate, but
order-to-order comparisons are less meaningful.
We have made no attempt to model the Swings effect for any molecule. 
Without such a model, any line could be a different strength than expected
from laboratory measurements.  Also, the lines are often highly blended.
Thus, to determine line strengths a full deconvolution model must
be computed.  For the over 4000 unidentified lines this is a huge
undertaking.  Such an exercise should be done for individual studies,
but are not realistic for this atlas.

In addition to the tables of identified and unidentified features,
we are making available PostScript files of plots of our spectra
in two formats.  The first format is plots similar to Fig.~\ref{order49}.
For each order, there are two plots, each with 3 panels, showing two
spectra and the identified and unidentified lines.  These are all
plotted with expanded y scales to show the weaker features.
The features are marked in color, as with Fig.~\ref{order49}.
The colors of the markers remain the same for each molecule across
all plots.  That is, there is space and a color allocated for each
molecule even when that molecule is not present in a particular order.
Included in these plots are the ionic species CH$^+$ and H$_{2}$O$^+$, since
they are visible in the optocenter spectra.
The CO$^+$ only shows up in order 81 in the tail spectrum.  This
spectrum is shown in Fig.~2 of Cochran {\it et al.} (2000)
with the CO$^+$ and CH$^+$ identified.  

The second form of plot which we are making available is 
the spectra without expansion of the y axis.  These plots
take the form of four panel plots with two orders per plot (each of
two panels).  Each panel is scaled in y to fit the maximum
strength line in that panel.  These plots are a valuable adjunct to the
identification plots because they show the relative strength of the strong
and weak lines.  They also allow for
the inspection of the telluric absorption features which plague some orders.

It is quite likely that there are a few real features which we
have missed in our tabulations because the sheer quantity of data
makes such misses inevitable.  Indeed, every time we have inspected
the plots of identifications, we have found an additional line or two.
There are some lines which we may have located which may be less
believable to a reader.  Sometimes our choice to include these lines
in our lists comes from our ability to manipulate the spectra digitally
and to change the scaling to view a feature better; sometimes we may have
overactive imaginations!  There are times when an identified molecular
line seems slightly off in wavelength from the peak of a feature.
We have assumed that it is possible that the wavelengths of the laboratory
line lists are not always entirely accurate.  Additionally, in regions
where lines are blended, or in the regions of strong telluric absorption
features, the observed line center may be shifted from a laboratory value.
This is true also for our measured positions of unidentified features.
Therefore, it is imperative that a user of these data inspect the
full scale plots for context and for absorptions and that they examine the
feature plots for blending.  Users who wish to model the unidentified
features should contact us for access to the data, so that the lines
can be properly deblended.

\section{Summary}
In 1995, we obtained high spectral resolution observations of the
dust-poor comet
122P/de~Vico at McDonald Observatory.  We have compiled many
laboratory studies of molecules thought to be present in the spectra of
comets and have used these to identify 12,219 lines in the spectrum of de~Vico.
In addition, we found 4,055 lines we could not attribute to a particular
molecule.  We found emission lines of five neutral radicals:
6862 were of C$_{2}$, 1282 of C$_{3}$,
1167 of CN, 169 of CH, and 2569 of NH$_{2}$.  We also identified three
atomic lines.  We observed three ionic species: 129 lines of H$_{2}$O$^+$,
11 lines of CH$^+$ and 27 lines of CO$^+$.  The CO$^+$ can only be
seen in the spectrum obtained 100\,arcsec tailward, while
the H$_{2}$O$^+$ and CH$^+$ can be seen in both the tail and optocenter
spectra.

Our lists of identified and unidentified emission lines are available
in digital format from the Planetary Data System Small Bodies Node.
In addition, we are making available annotated plots of high
resolution spectra of a comet, covering almost the complete optical
bandpass.  The line lists, coupled with the plots, should prove
to be an invaluable tool for future studies of high resolution cometary
spectra.  Only by first identifying the molecules which are
present in a spectrum can the important chemical analyses be performed.

\vspace{10pt}
\begin{center}Acknowledgments\end{center}
This work was funded by NASA Grant NAG5 9003.  We are indebted to
Dr. Claude Arpigny for encouragement and many helpful discussions.
We thank Dr. Eric Bakker for pointers to several of the line lists.
We thank Dr. Tony Farnham for a careful reading of the manuscript.
We are particularly indebted to the various people who made digital
line lists available, either at our requests or as part of a database
attached to a paper.  These include Drs. Balfour, Brown, Davis, Destombes,
Haridass, J{\o}rgensen, Kurucz, Osterbrock, and  Ross.

\newpage

\clearpage
\begin{table}
\caption{Log of Observations \label{log}}
\centering
\begin{tabular}{lccccccc}
\hline
\hline
 & & Start & \\
 & Date & Time & Exposure & $\Delta$ & {$\dot\Delta$} & R$_h$ & \.{R}$_h$ \\
\multicolumn{1}{c}{File \#} & (UT) & (UT) & (sec) & ({\sc au}) &
(km/sec) & ({\sc au}) & (km/sec) \\
\hline
RV23200 & 03 Oct 95 & 11:37:30 & 600 & 1.00 & -14.3 & 0.66 & -2.9 \\
RV23201 & 03 Oct 95 & 11:57:17 & 600 & 1.00 & -14.3 & 0.66 & -2.9 \\
RV23257 & 04 Oct 95 & 11:10:49 & 1500 & 0.99 & -12.9 & 0.66 & -1.7 \\
RV23258$^a$ & 04 Oct 95 & 11:42:21 & 1200 & 0.99 & -12.9 & 0.66 & -1.7 \\
\hline
\multicolumn{8}{l}{\mbox{}\hspace{0.25in}Note: a) 100 arcsec tailward} \\
\end{tabular}

\end{table}

\begin{table}
\caption{Wavelengths Observed \label{wave}}
\centering
\begin{tabular}{ccc|ccc|ccc}
\hline \hline
Order & Start & End & Order & Start & End & Order & Start & End \\
 & (\AA) & (\AA) & & (\AA) & (\AA) & & (\AA) & (\AA) \\
\hline
 89 &    3829.77 &    3893.62 &	 70 &    4869.76 &    4950.90 &	 51 &    6684.03 &    6795.33 \\
 88 &    3873.32 &    3937.89 &	 69 &    4940.35 &    5022.66 &	 50 &    6817.70 &    6931.23 \\
 87 &    3917.88 &    3983.19 &	 68 &    5013.01 &    5096.53 &	 49 &    6956.83 &    7072.67 \\
 86 &    3963.46 &    4029.53 &	 67 &    5087.84 &    5172.60 &	 48 &    7101.75 &    7220.00 \\
 85 &    4010.12 &    4076.96 &	 66 &    5164.94 &    5250.98 &	 47 &    7252.84 &    7373.60 \\
 84 &    4057.89 &    4125.53 &	 65 &    5244.41 &    5331.77 &	 46 &    7410.50 &    7533.88 \\
 83 &    4106.81 &    4175.26 &	 64 &    5326.36 &    5415.08 &	 45 &    7575.16 &    7701.29 \\
 82 &    4156.91 &    4226.20 &	 63 &    5410.91 &    5501.04 &	 44 &    7747.31 &    7876.30 \\
 81 &    4208.26 &    4278.40 &	 62 &    5498.18 &    5589.76 &	 43 &    7927.46 &    8059.45 \\
 80 &    4260.88 &    4331.90 &	 61 &    5588.32 &    5681.40 &	 42 &    8116.19 &    8251.32 \\
 79 &    4314.84 &    4386.75 &	 60 &    5681.46 &    5776.09 &	 41 &    8314.13 &    8452.55 \\
 78 &    4370.18 &    4443.01 &	 59 &    5777.75 &    5873.98 &	 40 &    8521.97 &    8663.85 \\
 77 &    4426.96 &    4500.73 &	 58 &    5877.37 &    5975.26 &	 39 &    8740.46 &    8885.98 \\
 76 &    4485.23 &    4559.97 &	 57 &    5980.48 &    6080.08 &	 38 &    8970.45 &    9119.79 \\
 75 &    4545.05 &    4620.78 &	 56 &    6087.27 &    6188.65 &	 37 &    9212.88 &    9366.25 \\
 74 &    4606.48 &    4683.24 &	 55 &    6197.94 &    6301.16 &	 36 &    9468.77 &    9626.40 \\
 73 &    4669.60 &    4747.41 &	 54 &    6312.71 &    6417.84 &	 35 &    9739.28 &    9901.42 \\
 72 &    4734.47 &    4813.35 &	 53 &    6431.82 &    6538.93 &	 34 &   10025.71 &   10192.61 \\
 71 &    4801.16 &    4881.16 &	 52 &    6555.50 &    6664.66 &	 & & \\
\hline
\end{tabular}
\end{table}

\begin{table}
\caption{The Lines of Order 49, panel 1}\label{lines}
\begin{tabular}{rrlccl}
\hline
 \\ [-9pt]
\multicolumn{1}{c}{Wavelength} & \multicolumn{1}{c}{Frequency} & &
Electronic$^1$ & Vibrational$^2$ & \multicolumn{1}{c}{Rotational$^3$} \\
\multicolumn{1}{c}{(\AA)} & \multicolumn{1}{c}{(cm$^{-1}$)} & 
\multicolumn{1}{c}{Molecule} & Transition & Transition & \multicolumn{1}{c}{Transitions} \\
\hline \hline
6956.831  &  14370.397  &  H2O+  &  A--X  &  (0,2,0) -- (0,0,0)  &  3 1 2 --  2 0 2 \\
6957.301  &  14369.430  &  CN  &  A--X  &  3--0  &  P1( 2) \\
6957.612  &  14368.780  &  CN  &  A--X  &  3--0  &  P2( 9) \\
6957.731  &  14368.540  &  CN  &  A--X  &  3--0  &  Q1( 8)  R1(19) \\
6958.122  &  14367.730  &  NH2  &  A--X  &  (0, 2,0) -- (0,0,0)  &  5 1 5 --  5 0 5 \\
6958.228  &  14367.513  &  H2O+  &  A--X  &  (0,2,0) -- (0,0,0)  &  3 1 2 --  2 0 2 \\
6958.640  &  14366.655  &  Unid  &    &    &   \\
6958.864  &  14366.200  &  C2  &  Phil  &  4--0  &  Q(18) \\
6958.907  &  14366.110  &  CN  &  A--X  &  3--0  &  P1( 3) \\
6959.100  &  14365.710  &  CN  &  A--X  &  3--0  &  Q2(14) \\
6959.394  &  14365.100  &  CN  &  A--X  &  3--0  &  Q1( 9) \\
6960.547  &  14362.730  &  CN  &  A--X  &  3--0  &  R1(20) \\
6960.805  &  14362.190  &  CN  &  A--X  &  3--0  &  P1( 4) \\
6960.850  &  14362.100  &  NH2  &  A--X  &  (0, 2,0) -- (0,0,0)  &  5 1 5 --  5 0 5 \\
6961.318  &  14361.130  &  CN  &  A--X  &  3--0  &  Q1(10) \\
6961.720  &  14360.312  &  Unid  &    &    &   \\
6961.903  &  14359.930  &  CN  &  A--X  &  3--0  &  P2(10) \\
6962.377  &  14358.950  &  C2  &  Phil  &  4--0  &  P(14) \\
6962.795  &  14358.090  &  CN  &  A--X  &  3--0  &  P12( 3) Q2(15) \\
6962.955  &  14357.760  &  CN  &  A--X  &  3--0  &  P1( 5) \\
6963.331  &  14356.983  &  H2O+  &  A--X  &  (0,2,0) -- (0,0,0)  &  2 1 1 --  1 0 1 \\
6963.496  &  14356.640  &  CN  &  A--X  &  3--0  &  Q1(11) \\
6963.569  &  14356.490  &  CN  &  A--X  &  3--0  &  R1(21) \\
6963.594  &  14356.440  &  NH2  &  A--X  &  (0, 2,0) -- (0,0,0)  &  4 1 4 --  4 0 4 \\
6963.880  &  14355.862  &  Unid  &    &    &   \\
6964.080  &  14355.447  &  Unid  &    &    &   \\
6964.303  &  14354.980  &  NH2  &  A--X  &  (0, 2,0) -- (0,0,0)  &  1 1 1 --  1 0 1 \\
6964.590  &  14354.394  &  Unid  &    &    &   \\
6964.800  &  14353.961  &  Unid  &    &    &   \\
6965.392  &  14352.740  &  CN  &  A--X  &  3--0  &  P1( 6) \\
6965.470  &  14352.575  &  H2O+  &  A--X  &  (0,2,0) -- (0,0,0)  &  2 1 1 --  1 0 1 \\
6965.901  &  14351.690  &  CN  &  A--X  &  3--0  &  Q1(12) \\
6966.064  &  14351.350  &  NH2  &  A--X  &  (0, 2,0) -- (0,0,0)  &  5 1 4 --  4 2 2 \\
6966.200  &  14351.070  &  C2  &  Phil  &  4--0  &  R(26) \\
6966.212  &  14351.050  &  CN  &  A--X  &  3--0  &  P12( 4) \\
6966.326  &  14350.810  &  NH2  &  A--X  &  (0, 2,0) -- (0,0,0)  &  4 1 4 --  4 0 4 \\
6966.452  &  14350.550  &  CN  &  A--X  &  3--0  &  P2(11) \\
6966.637  &  14350.170  &  NH2  &  A--X  &  (0, 2,0) -- (0,0,0)  &  2 1 2 --  2 0 2 \\
6966.700  &  14350.040  &  NH2  &  A--X  &  (0, 2,0) -- (0,0,0)  &  3 1 3 --  3 0 3 \\
6966.719  &  14350.000  &  CN  &  A--X  &  3--0  &  Q2(16) \\
6967.060  &  14349.305  &  Unid  &    &    &   \\
\hline
\end{tabular}
\end{table}

\begin{table}
\begin{center}{Table~\ref{lines} (cont.)}\end{center}
\begin{tabular}{rrlccl}
\hline
 \\ [-9pt]
\multicolumn{1}{c}{Wavelength} & \multicolumn{1}{c}{Frequency} & &
Electronic$^1$ & Vibrational$^2$ & \multicolumn{1}{c}{Rotational$^3$} \\
\multicolumn{1}{c}{(\AA)} & \multicolumn{1}{c}{(cm$^{-1}$)} &
\multicolumn{1}{c}{Molecule} & Transition & Transition & \multicolumn{1}{c}{Transitions} \\
\hline \hline
6967.380  &  14348.646  &  Unid  &    &    &   \\
6967.730  &  14347.909  &  Unid  &    &    &   \\
6967.904  &  14347.560  &  NH2  &  A--X  &  (0, 2,0) -- (0,0,0)  &  5 1 4 --  4 2 2 \\
6968.061  &  14347.240  &  CN  &  A--X  &  3--0  &  P1( 7) \\
6968.260  &  14346.832  &  Unid  &    &    &   \\
6968.584  &  14346.160  &  CN  &  A--X  &  3--0  &  Q1(13) \\
6968.760  &  14345.798  &  Unid  &    &    &   \\
6969.124  &  14345.050  &  C2  &  Phil  &  4--0  &  Q(20) \\
6969.396  &  14344.490  &  NH2  &  A--X  &  (0, 2,0) -- (0,0,0)  &  3 1 2 --  2 2 0 \\
6969.993  &  14343.260  &  NH2  &  A--X  &  (0, 2,0) -- (0,0,0)  &  3 1 3 --  3 0 3 \\
6970.001  &  14343.240  &  CN  &  A--X  &  3--0  &  P12( 5) \\
6970.152  &  14342.934  &  H2O+  &  A--X  &  (0,2,0) -- (0,0,0)  &  1 1 0 --  0 0 0 \\
6970.499  &  14342.220  &  NH2  &  A--X  &  (0, 2,0) -- (0,0,0)  &  1 1 1 --  1 0 1 \\
6970.846  &  14341.500  &  CN  &  A--X  &  3--0  &  Q2(17) \\
6970.892  &  14341.410  &  NH2  &  A--X  &  (0, 2,0) -- (0,0,0)  &  2 1 2 --  2 0 2 \\
6971.032  &  14341.120  &  CN  &  A--X  &  3--0  &  P1( 8) \\
6971.181  &  14340.820  &  CN  &  A--X  &  3--0  &  P2(12) \\
6971.513  &  14340.130  &  CN  &  A--X  &  3--0  &  Q1(14) \\
6971.900  &  14339.341  &  Unid  &    &    &   \\
6972.137  &  14338.850  &  NH2  &  A--X  &  (0, 2,0) -- (0,0,0)  &  3 1 2 --  2 2 0 \\
6972.600  &  14337.908  &  Unid  &    &    &   \\
6973.518  &  14336.010  &  C2  &  Phil  &  4--0  &  P(16) \\
6973.720  &  14335.596  &  H2O+  &  A--X  &  (0,2,0) -- (0,0,0)  &  1 1 0 --  0 0 0 \\
6973.962  &  14335.100  &  CN  &  A--X  &  3--0  &  P12( 6) \\
6974.248  &  14334.510  &  CN  &  A--X  &  3--0  &  P1( 9) \\
6974.693  &  14333.600  &  CN  &  A--X  &  3--0  &  Q1(15) \\
6975.193  &  14332.570  &  CN  &  A--X  &  3--0  &  Q2(18) \\
6976.128  &  14330.650  &  CN  &  A--X  &  3--0  &  P2(13) \\
\hline
NOTES: \\
\multicolumn{1}{r}{1:} & \multicolumn{5}{p{12.5cm}}{For C$_{2}$, ``Phil" is the Phillips band (A~$^1\Pi_u$~--~X~$^1\Sigma^+_g$).} \\
\multicolumn{1}{r}{2:} & \multicolumn{5}{p{12.5cm}}{The designation of the vibrational
transition is different for diatomic and polyatomic molecules.} \\
\multicolumn{1}{r}{3:} & \multicolumn{5}{p{12.5cm}}{The rotational transition designations depend on the molecule.} \\
&   \multicolumn{5}{p{12.5cm}}{For CN, more than one transition may be listed per line
if they occur at the same wavelength. } \\
& \multicolumn{5}{p{12.5cm}}{For NH$_{2}$, rotational quantum numbers are
listed as $N^{\prime}_{K^{\prime}_{a}K^{\prime}_{c}} - N^{\prime\prime}_{K^{\prime\prime}_{a}K^{\prime\prime}_{c}}$.} \\
\\ [-9pt]
\hline
\end{tabular}
\end{table}

\newpage
\clearpage
\section{Figure Captions}

\begin{figure}[p]
\vspace{8.0in}
\includegraphics{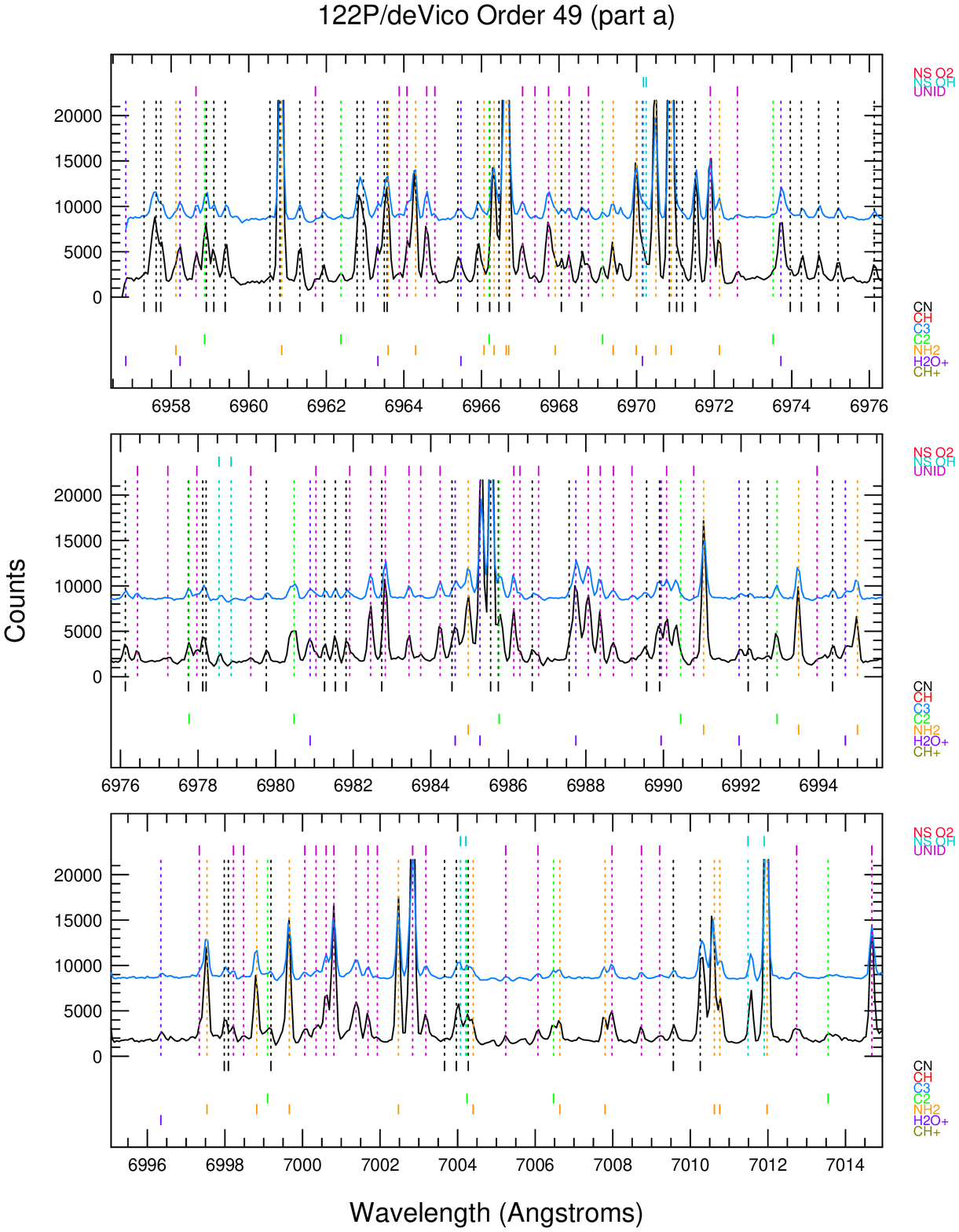}
\caption[fig1]{Cochran and Cochran 2001}\label{order49}
\end{figure}

\noindent
{\bf Figure~\ref{order49}:}
A representative spectral order is shown with an expanded y scale in order
to demonstrate the detection of weak features.  Two spectra are plotted.
The bottom, black spectrum, is RV23257, from 4 October 1995.
The top, blue spectrum, is RV23201, from 3 October 1995.  The blue
spectrum is offset vertically for clarity.  Marked on the plot are all of
the identified and unidentified features in the spectrum.

\begin{figure}[p]
\vspace{8.0in}
\includegraphics{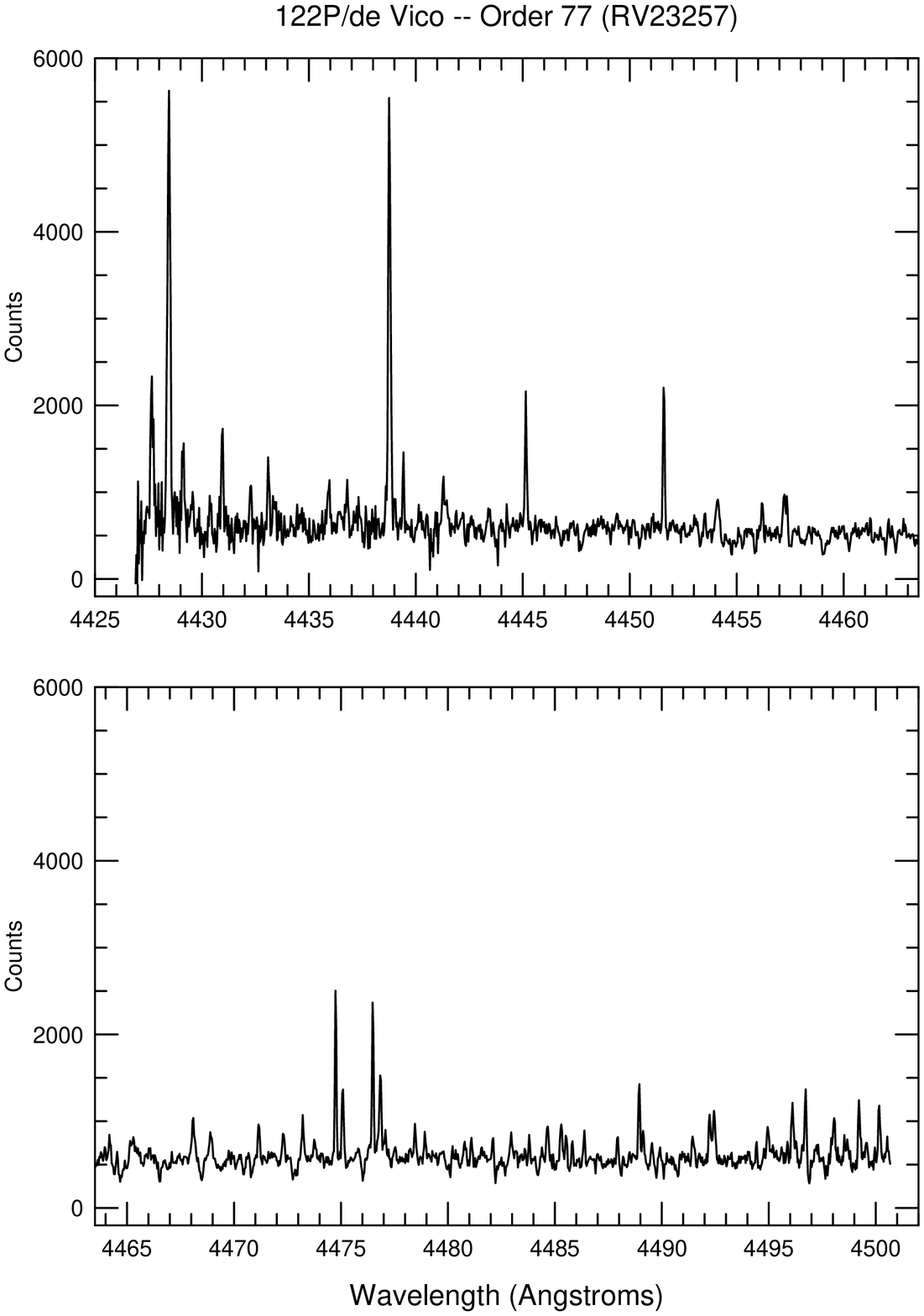}
\caption[fig1]{Cochran and Cochran 2001}\label{order77}
\end{figure}

\noindent
{\bf Figure~\ref{order77}:}
Spectral order 77 shows many strong emission features but we can not identify
the molecule responsible for any of these lines.  There are 64 emission lines in
this spectral region.

\begin{figure}[p]
\vspace{7in}
\includegraphics{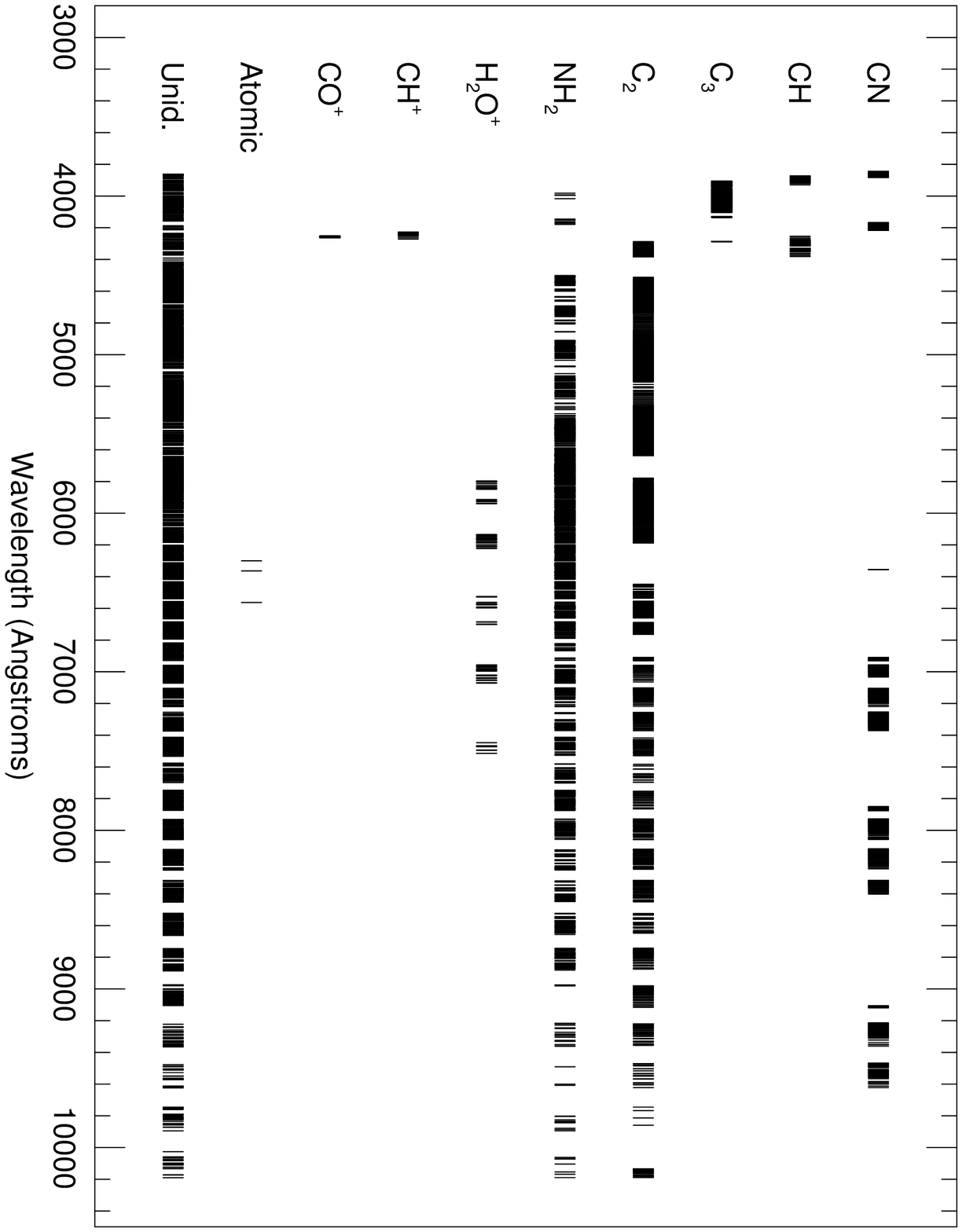}
\caption[fig1]{Cochran and Cochran 2001}\label{wavel}
\end{figure}

\noindent
{\bf Figure~\ref{wavel}:}
The wavelengths of the detected lines of the various molecules and atoms
we detected are plotted. 
The gaps at the red end for molecules such as C$_{2}$ are
caused by the interorder gaps in our spectra.  The wavelengths where
we have unidentified lines are also marked.

\clearpage

\end{document}